\documentclass{aa}
\usepackage{graphics}
\def\pdot{ {\dot {P}} }
\def\erot{ {{L}}_{\rm sd}}
\def\lx{L_{\rm x}}
\def\nh{N_{\rm H}}
\def\lxnor{L_{\rm x,nor}}
\def\lxcrit{L_{\rm x,crit}}
\def\gammacrit{L_{\rm \gamma,crit}}

\def\lxasca{L_{\rm x, (2-10)}}
\def\ergs{ {\rm {erg \,\,s}^{-1}} }
\begin{document}
\title{Re-examining the X-Ray versus Spin-Down Luminosity Correlation of 
Rotation Powered Pulsars}
\author{
Andrea Possenti \inst{1}
\and
Rossella Cerutti \inst{2}
\and
Monica Colpi \inst{3}
\and
Sandro Mereghetti \inst{4}        
}
\offprints{A. Possenti}
\institute{
Osservatorio Astronomico di Bologna, 
via Ranzani 1, I-40127 Bologna, Italy\\
email: phd@tucanae.bo.astro.it
\and
Dipartimento di Fisica, Universit\`a di Milano ,
via Celoria 16, I-20133 Milano, Italy \\
e-mail: rossella@ifctr.mi.cnr.it
\and
Dipartimento di Fisica, Universit\`a di Milano Bicocca,
Piazza della Scienza 3, I-20126 Milano, Italy \\
email: colpi@castore.mib.infn.it
\and
Istituto di Astrofisica Spaziale e Fisica Cosmica, 
Sezione di Milano ``G.Occhialini'', CNR, 
via Bassini 15, I-20133 Milano, Italy\\
email: sandro@ifctr.mi.cnr.it 
}
\date{Accepted for publication by {\sl Astronomy \& Astrophysics}}

\authorrunning{Possenti et al.}
\titlerunning{Re-examining the $\lx$ versus $\erot$ correlation of Pulsars}

\abstract{The empirical relation between the X-ray luminosity (in the
2-10 keV band) and the rate of spin-down energy loss $\erot$ of a
sample of 39 pulsars is re-examined considering recent data from {\sl
ASCA, RXTE, BeppoSAX, Chandra}, and {\sl XMM-Newton} and including
statistical and systematic errors.  The data show a significant
scatter around an average correlation between $\lxasca$ and $\erot$.
By fitting a dependence of $\lxasca$ on the period $P$ and period
derivative $\dot P$ of the type $\lxasca\propto P^a\,\pdot^b$, we
obtain $a=-4.00$ and $b=+1.34$ (i.e. $a\simeq -3b$).  This translates
into the relation $\lxasca=\lxnor \left ({\erot/ \ergs}\right
)^{1.34}$ with a normalization $\lxnor=10^{-15.3}~\ergs$. However, the
reduced $\chi^2$ is large ($=7.2$) making the fit unacceptable on
statistical ground.  All the X-ray luminosities lie below a critical
line $\lxcrit:$ the corresponding efficiency of conversion of
rotational energy into 2-10 keV X-rays is $\eta_{\rm
x}=(\lxcrit/\erot)\propto \erot^{0.48}$ and varies, within the sample,
between 0.1 and 80\%.  The large dispersion of $\lx$ below $\lxcrit$
indicates that other physical parameters uncorrelated with $P$ and
$\pdot$ need to be included to account for the observed emission at
X-ray energies.  We indicate a few possibilities that all conspire to
reduce $\lxasca.$ }

\maketitle 

\section{Introduction}
About 40 of the nearly 1,400 known radio pulsars (Manchester 2001)
have been so far detected in the X-ray range.  Despite the great
differences in their rotational parameters, magnetic field and age,
these rotation-powered X-ray sources seem to obey a simple empirical
law. The data reveal a correlation between their high-energy
luminosity and their spin-down luminosity $\erot=4\pi^2 I \pdot/P^3$,
where $P$ and $\pdot$ are the spin period and its derivative, and $I$
is the neutron star momentum of inertia that we assume 
equal to 10$^{45}$ g cm$^2$.

This correlation was first noticed in a small set of radio pulsars by
Seward and Wang (1988), and later investigated by Becker and
Tr\"{u}mper (1997; BT97 hereafter) using a sample of 27 powered
pulsars detected with {\sl ROSAT} in the soft energy band (0.1-2.4
\,\,\rm {keV}).  BT97 described the available data with a simple
scaling relation
\begin{equation}
\lx\simeq 10^{-3}\erot \qquad  . 
\end{equation}
However, a harder energy interval seems preferred to explore the
nature of the X-ray emission resulting from the rotational energy
loss: at energies above $\sim$2 keV the contribution from the neutron
star cooling and the spectral fitting uncertainties due to
interstellar absorption are reduced.  Saito (1998) examined the
correlation between $\erot$ and the pulsed X--ray luminosity using a
sample of radio pulsars detected at high energies with {\sl ASCA}
(2-10 keV) finding the scaling relation
\begin{equation} 
\lxasca\simeq10^{-21} \erot^{3/2} \qquad  . 
\end{equation}
 
None of these studies included the statistical and systematic
uncertainties on $\lx$ resulting from the errors on the X-ray fluxes
and on the poorly known distances of most pulsars.  A proper treatment
of these uncertainties is required to quantify the strength of the
correlation.

The physical origin of the non-thermal X-ray emission from radio
pulsars is uncertain. Polar cap models give predictions on the pulsed
X-ray luminosity, which is attributed to inverse Compton scattering of
higher order generation pairs on soft photons emitted by the surface
of the neutron star and/or by hot polar caps (Zhang \& Harding 2000).
The soft tail in the inverse Compton scattering spectrum can explain
the non-thermal X-ray component observed in many pulsars.

Outer gap models attribute the pulsed non-thermal X-ray emission to
synchrotron radiation of downward cascades from the outer gap
particles, and include (as in the inner gap models) a thermal
component from the hot polar caps heated by impinging particles.  Both
models are consistent with the scaling relation $\lx\propto \erot$
(Cheng, Gil, \& Zhang 1998) in the {\sl ROSAT} domain, but are in
conflict with the observations in the {\sl ASCA} (Zhang \& Harding
2000) domain where they over or under-predict the values of $\lx$
(relative to relation [2]).  In polar cap models, these discrepancies
can be accounted for by invoking {\it ad hoc} model parameters for the
single sources, due to the difficulty in tracking the full cascade of
particles and photons down to X-ray energies, coupled with the
uncertainties in the temperature of the polar cap and of the whole
surface.

The correlation between $\lx$ and $\erot$ is even more difficult to
quantify when considering that pulsed emission is observed only in 15
of the sources.  Thus, for most of the pulsars, the total luminosity
$\lx$ can be considered only as an upper limit on the pulsed
component.  $\lx$ often contains also the (unpulsed) contribution
resulting from pulsar wind nebular emission (Chevalier 2000) which is
in end related to the spin parameters $P$ and $\dot P$.
 
For a comparison of the theoretical predictions with the observations,
a statistical analysis of the data seems useful. Given the improved
capabilities of the {\sl Chandra} and {\sl XMM-Newton} observatories,
it is now timely to reconsider the correlation $\lx$ versus $\erot$ in
statistical terms, and this is the aim of our paper.

In $\S 2$ we describe the current sample and the criteria used to
account for the uncertainties in $\lx$.  In $\S 3$ we present the
results. In $\S 4$ we comment on the possible origin of the large
scatter seen in the data and in $\S 5$ we summarize our conclusions.

\begin{table*}[tbh]
\begin{center}
\caption{Parameters and emission of the 41 sample pulsars}
\label{tab:fit}
\vspace{0.0cm}
\tabcolsep 0.08truecm
\scriptsize{
\begin{tabular}{|l|c|c|c|c|c|c|c|c|c|c|} 
\hline
{\em ~~~Pulsar}&{\em P}& {\em $\dot P$}    &    {\em $N_H$}      &     {\em distance}     &{\em $\alpha$~{\rm or}~kT}&  {\em Obs Band}  & {\em Detector}  
&{\em $f^{(2-10)}$}          & {\em $\log L_X^{(2-10)}$}& ref\\
    ~~~~  PSR &  ms  &  $10^{-15}~~{\rm s\;s^{-1}}$ &$\rm 10^{21} \; cm^{-2}$ &       kpc  & \rm  ~~~~~ keV           &    \rm keV       &      
& $\rm erg \;s^{-1}cm^{-2}$    &  $\rm erg \;s^{-1}$ &  \\ \hline \hline 
                
J0030+0451  &   4.87 & $1.00\cdot 10^{-5}$ & $2.15\pm 0.85$      & $0.230\pm 0.092$          & $2\pm 0.2$        &  0.1--2.4  &  PSPC    
& $(1.27_{-0.61}^{+0.95})\cdot  10^{-13}$   & $29.88_{-0.73}^{+0.54} $ & 1\\               
J0218+4232  &   2.32 & $7.50\cdot 10^{-5}$ & $2\pm 2$            & $5.70 \pm 2.28$           & $0.94\pm 0.22$    &  2.0--10.0 &  MECS     
& $(4.30_{-0.22}^{+0.28})\cdot  10^{-13}$   & $33.20_{-0.47}^{+0.32} $ & 2\\
J0437--4715 &   5.76 & $1.86\cdot 10^{-5}$ & $0.8_{-0.6}^{+1}$   & $0.178\pm 0.026$          & $2.35\pm 0.35$    &  0.1--2.4  &  HRI      
& $(4.30_{-1}^{+1})\cdot        10^{-13}$   & $30.19_{-0.25}^{+0.21} $ & 3,4\\
J0751+1807  &   3.48 & $8.00\cdot 10^{-6}$ & $4.4_{-0.4}^{+4.6}$ & $2\pm 0.8$                & $2\pm 0.2$        &  0.1--2.4  &  PSPC    
& $(4.29_{-1.44}^{+3.54})\cdot  10^{-14}$   & $31.29_{-0.62}^{+0.55} $ & 3\\
J1012+5307  &   5.26 & $1.46\cdot 10^{-5}$ & $0.06\pm 0.01$      & $0.520\pm 0.208$          & $2.3\pm 0.2$      &  0.1--2.4  &  PSPC    
& $(1.25_{-0.61}^{+0.89})\cdot  10^{-14}$   & $29.58_{-0.73}^{+0.53} $ & 3\\
J1024--0719 &   5.16 & $2.99\cdot 10^{-6}$ & $0.2\pm 0.05$       & $0.350\pm 0.140$          & $2\pm 0.2$        &  0.1--2.4  &  HRI    
& $(8.86_{-3.36}^{+4.70})\cdot  10^{-15}$   & $29.09_{-0.79}^{+0.50} $ & 3\\
J1744--1134 &   4.07 & $7.13\cdot 10^{-6}$ & $0.1\pm 0.05$       & $0.357_{-0.035}^{+0.043}$ & $2\pm 0.2$        &  0.1--2.4  &  HRI    
& $(6.44_{-2.86}^{+4.08})\cdot  10^{-15}$   & $28.97_{-0.40}^{+0.32} $ & 3\\
B1821--24   &   3.05 & $1.62\cdot 10^{-3}$ & $2.9\pm 2.3$        & $5.1\pm 0.5$              & $1.89\pm 0.21$    &  0.7--10.0 &  GIS    
& $(1.25_{-0.69}^{+0.33})\cdot  10^{-12}$   & $33.56_{-0.44}^{+0.18} $ & 5\\
B1937+21    &   1.56 & $1.06\cdot 10^{-4}$ & $21\pm 5$           & $3.60\pm 1.44$         & $1.71_{-0.08}^{+0.05}$ & 0.5--10.0  & LECS+MECS  
& $(3.70_{-0.40}^{+0.40})\cdot  10^{-13}$   & $32.73_{-0.55}^{+0.39} $ & 6\\
J2124--3358 &   4.93 & $1.30\cdot 10^{-5}$ & $0.35\pm 0.15$      & $0.25\pm 0.10$            & $2\pm 0.2$        &  0.1--2.4  &  HRI    
& $(8.26_{-3.48}^{+0.45})\cdot  10^{-14}$   & $29.77_{-0.68}^{+0.32} $ & 7\\\hline
B0950+08    & 253.07 & $0.229            $ &                     & $0.127\pm 0.013$          &                   &  0.1--2.4  &  PSPC  
& $(2.3_{-0.7}^{+0.7})\cdot     10^{-14}$   & $28.62_{-0.62}^{+0.42} $ & 3\\
B1929+10    & 226.52 & $1.16             $ & $0.1\pm 0.05$       & $0.25\pm 0.08$            & $0.44\pm 0.046\;^c$ &  0.5--5.0  &  SIS  
& $(5.6_{-1.4}^{+1.5})\cdot     10^{-14}$   & $29.60_{-0.46}^{+0.34} $ & 8\\ 
B0823+26    & 530.66 & $1.71             $ &                     & $0.380\pm 0.152$          &                   &  0.1--2.4  &  PSPC   
& $(0.6_{-0.2}^{+0.2})\cdot     10^{-14}$   & $28.99_{-0.62}^{+0.42} $ & 3 \\\hline
B0114+58    & 101.44 & $5.85             $ & $2.57\pm 0.2$       & $2.14\pm 0.856$           & $2.1\pm 0.2$      &  0.1--2.4  &  PSPC   
& $(4.25_{-0.??}^{+4.25})\cdot  10^{-15}$   & $30.34_{-0.??}^{+0.59} $ & 9\\
B0355+54    & 156.38 & $4.40             $ & $0.2\pm 0.2$        & $2.10\pm 0.84$            & $2\pm 0.5$        &  0.1--2.4  &  PSPC   
& $(1.16_{-0.94}^{+3.04})\cdot  10^{-13}$   & $31.76_{-1.17}^{+0.85} $ & 10\\
J0538+2817  & 143.16 & $3.67             $ & $0.6\pm 0.6$        & $1.5\pm 0.6$              & $1.5\pm 0.5$      &  0.1--2.4  &  PSPC   
& $(8.00_{-0.??}^{+8.00})\cdot  10^{-16}$   & $29.31_{-0.??}^{+0.59} $ & 11\\
B0633+17    & 237.09 & $11.0             $ & $0.13\pm 0.13$      & $0.154_{-0.034}^{+0.059}$ & $2.19\pm 0.35$    &  0.7--5.0  &  GIS   
& $(7.94_{-2.2}^{+3.0})\cdot    10^{-14}$   & $29.33_{-0.36}^{+0.42} $ & 12\\
B0656+14    & 384.89 & $55.0             $ & $0.17\pm 0.17$      & $0.28_{-0.10}^{+0.20}$    & $1.5\pm 1.1$      &  1.0--5.0  &  SIS    
& $(2.05_{-0.74}^{+1.72})\cdot  10^{-13}$   & $30.26_{-0.58}^{+0.73} $ & 13\\
B1055--52   & 197.11 & $5.83             $ & $0.26\pm 0.06$      & $0.5\pm 0.2$              & $1.5\pm 0.3$      &  2.0--10.0 &  GIS    
& $(1.06_{-0.09}^{+0.10})\cdot  10^{-14}$   & $29.48_{-0.48}^{+0.33} $ & 14\\          
B1951+32    &  39.53 & $5.84             $ & $3.4\pm 0.5$        & $2.5\pm 0.2$              & $2.1\pm 0.3$      &  0.1--2.4  &  PSPC   
& $(2.04_{-0.79}^{+0.85})\cdot  10^{-12}$   & $33.16_{-0.28}^{+0.22} $ & 15\\\hline
B0833--45   &  89.33 & $1.25\cdot 10^{2} $ & $0.4\pm 0.1$        & $0.25\pm 0.03$            & $2.2\pm 0.4$      &  0.2--8.0  &  ACIS-S 
& $(1.03_{-0.55}^{+1.04})\cdot  10^{-11}$   & $31.86_{-0.44}^{+0.40} $ & 16\\
B1046--58   & 123.67 & $96.3             $ & $5\pm 1 $           & $2.98\pm 1.19$            & $2\pm 0.2$        &  0.4--10.0 &  SIS    
& $(2.50_{-0.58}^{+0.66})\cdot  10^{-13}$   & $32.40_{-0.56}^{+0.39} $ & 17\\
J1105--6107 &  63.19 & $15.8             $ & $8.55\pm 5.25$      & $7.0\pm 2.8$              & $1.8\pm 0.4$      &  2.0--10.0 &  GIS    
& $(6.47_{-1.04}^{+1.18})\cdot  10^{-13}$   & $33.55_{-0.52}^{+0.37} $ & 18\\
J1420--6048 &  68.18 & $83.2             $ & $22\pm 7$           & $2.0\pm 0.8$              & $1.6\pm 0.4\;^b$  &  2.0--10.0 &  GIS    
& $(4.70_{-0.74}^{+0.77})\cdot  10^{-12}$   & $33.33_{-0.52}^{+0.36} $ & 19\\
B1706--44   & 102.46 & $93.0             $ & $3.45\pm 3.45$      & $1.80\pm 0.72$            & $1.9\pm 0.9$      &  2.0--10.0 &  SIS+GIS
& $(1.03_{-0.24}^{+0.38})\cdot  10^{-12}$   & $32.58_{-0.56}^{+0.43} $ & 20\\
B1757--24   & 124.90 & $1.28\cdot 10^{2} $ & $35\pm 12$          & $5.0_{-0.7}^{+2.0}$       & $1.6\pm 0.6$      &  2.0--10.0 &  ACIS-S 
& $(7.9_{-0.6}^{+0.6})   \cdot  10^{-13}$   & $33.37_{-0.10}^{+0.20} $ & 21\\
B1800--21   & 133.63 & $1.34\cdot 10^{2} $ & $13\pm 1$           & $5.30\pm 2.12$            & $2\pm 0.2$        &  0.1--2.4  &  PSPC   
& $(1.78_{-0.59}^{+0.70})\cdot  10^{-13}$   & $32.75_{-0.70}^{+0.45} $ & 22\\
J1811--1926 &  64.67 & $44.0             $ & $13.8\pm 0.8$       & $7.8\pm 2.5$              & $1.89\pm 0.25$    &  4.0--10.0 &  MECS   
& $(1.23_{-0.11}^{+0.07})\cdot  10^{-11}$   & $34.93_{-0.38}^{+0.27} $ & 23\\
B1823--13   & 101.45 & $75.5             $ & $40\pm 25$          & $4.12\pm 1.65$            & $2\pm 0.2$        &  0.5--2.4  &  PSPC   
& $(1.70_{-1.4}^{+4.4})\cdot    10^{-11}$   & $34.51_{-1.20}^{+0.85} $ & 24\\
B1853+01    & 267.40 & $2.08\cdot 10^{2} $ & $2.57\pm 0.2$       & $3.2\pm 1.3$              & $2.3\pm 1.1$      &  0.4--2.0  &  GIS    
& $(1.2_{-0.3}^{+0.3})\cdot     10^{-12}$   & $33.14_{-0.57}^{+0.39} $ & 25\\
J2229+6114  &  51.62 & $78.0             $ & $6.3\pm 1.3$        & $3\pm 1$                  & $1.51\pm 0.14$    &  2.0--10.0 &  ACIS-I 
& $(1.30_{-0.08}^{+0.09})\cdot  10^{-12}$   & $33.12_{-0.38}^{+0.28} $ & 26\\
B2334+61    & 495.28 & $1.92\cdot 10^{2} $ & $2\pm 1$            & $2.5\pm 1$                & $2\pm 0.2$        &  0.1--2.4  &  PSPC   
& $(4.05_{-1.7}^{+2.6})\cdot    10^{-14}$   & $31.46_{-0.86}^{+0.52} $ & 27\\\hline
J0205+6449  &  65.68 & $1.93\cdot 10^{2} $ & $3\pm 2$            & $2.6 \pm 0.6$             & $1.9 \pm 0.2\;^b$ &  0.8--10.0 &  HRC    
& $(1.5_{-0.3}^{+0.3})\cdot     10^{-11}$   & $34.08_{-0.21}^{+0.14} $ & 28\\
B0531+21    &  33.52 & $4.21\cdot 10^{2} $ & $3\pm 0.5$          & $2\pm 0.5$                & $2.108\pm 0.006$  &  0.3--10.0 &  MOS   
& $(9.93_{-0.43}^{+0.09})\cdot  10^{-9} $   & $36.65_{-0.27}^{+0.20} $ & 29\\
J0537--6910 &  16.11 & $51.0             $ & $6.9\pm 3.3$        & $47.3\pm 0.8$             & \begin{tabular}{c} 
                                                                                               $1.6\pm 0.2\;^a$\\\hline 
                                                                                               $2.55\pm 0.15\;^b$\\
                                                                                               \end{tabular}      
                                                                                                                  &  0.2--10.0 &  GIS   
& $(5.13_{-1.37}^{+1.38})\cdot  10^{-12}$   & $36.11_{-0.15}^{+0.12} $ & 30\\             
B0540--69   &  50.53 & $4.73\cdot 10^{2} $ & $4.6\pm 4.6$        & $47.3\pm 0.8$             & \begin{tabular}{c} 
                                                                                               $1.83\pm 0.13\;^a$\\\hline
                                                                                               $2.06\pm 0.2\;^b$\\
                                                                                               \end{tabular} 
                                                                                                                  &  0.2--10.0 &  ACIS-I
& $(3.33_{-1.29}^{+0.97})\cdot  10^{-11}$   & $36.93_{-0.23}^{+0.13} $ & 31\\
J1119--6127 & 407.75 & $4.02\cdot 10^{3} $ & $15\pm 15$          & $5\pm 3$                  & $1.4_{-1.2}^{+1.0}$ &  0.7--5.0  &  GIS   
& $(4.74_{-2.7}^{+6.8})\cdot    10^{-13}$   & $33.13_{-1.16}^{+0.80} $ & 32\\ 
J1124--5916 & 135.31 & $7.45\cdot 10^{2} $ & $3.17\pm 0.15$      & $4.8 \pm 1.6$             & $1.9 \pm 0.2$      &  2.0--8.0  &  ACIS-S
& $(1.1_{-0.2}^{+0.2})\cdot     10^{-11}$   & $34.48_{-0.31}^{+0.18} $ & 33,34\\          
B1509--58   & 150.66 & $1.54\cdot 10^{3 }$ & $12.7\pm 12.7$      & $4.2\pm 0.5$              & \begin{tabular}{c} 
                                                                                               $1.358\pm 0.014\;^a$\\\hline
                                                                                               $2.2\pm 0.005\;^b$\\
                                                                                               \end{tabular} 
                                                                                                                  &  2.0--250  &  PCA   
& $(1.05_{-0.33}^{+0.06})\cdot  10^{-10}$   & $35.32_{-0.27}^{+0.12} $ & 35\\
J1617--5055 &  69.36 & $1.37\cdot 10^{2} $ & $6.8\pm 6.8$        & $4.5\pm 0.9$              & $1.6\pm 0.3$       &  3.5--10.0 &  GIS   
& $(8.86_{-0.34}^{+0.49})\cdot  10^{-12}$   & $34.31_{-0.21}^{+0.18} $ & 36\\             
J1846--0258 & 323.60 & $7.10\cdot 10^{3} $ & $47 \pm 8  $        & $19\pm 5$                 & $2.2 \pm 0.1$      &  3.0--20.0 &  PCA   
& $(3.90_{-0.4}^{+0.4})\cdot    10^{-11}$   & $36.22_{-0.32}^{+0.28} $ & 37,38\\
\hline
\end{tabular}
}
\end{center}
\vspace{0.0truecm}
\scriptsize{{\bf Refs}: 
{\underline{Note}}: $^a$ Pulsed component photon index; $^b$ compact nebula photon index;
$^c$ polar cap black body temperature.
{\underline{Detector}}: PSPC and HRI are instruments aboard of {\it ROSAT}; GIS and SIS are intruments 
aboard of {\it ASCA}; LECS, MECS and PCA are instruments aboard of {\it Beppo-SAX}; ACIS-I, ACIS-S and HRC 
are instruments aboard of {\it Chandra}; MOS is an instrument aboard of {\it Newton-XMM}.
{\underline{Obs Band}}: is the energy band of the observation from which the fluxes
reported in column 9 are measured or extrapolated.
{\underline{References}}: 
[1]  Becker et al. 2000; 
[2]  Mineo et al. 2000; 
[3]  BT99; 
[4]  Kawai et al. 1998;
[5]  Saito et al. 1997;
[6]  Nicastro et al. 2002; 
[7]  Sakurai et al. 2001;
[8]  Wang, Halpern 1997; 
[9]  Slane 1995;
[10] Slane 1994; 
[11] Sun et al. 1995;  
[12] Halpern, Wang 1997; 
[13] Greiveldinger et al. 1996; 
[14] Cheng, Zhang 1999;
[15] Chang, Ho 1997; 
[16] Pavlov et al. 2001; 
[17] Pivovaroff et al. 2000;
[18] Gotthelf, Kaspi 1998; 
[19] Roberts et al. 2001; 
[20] Finley et al. 1998;
[21] Kaspi et al. 2001;
[22] Finley, Ogelman 1994; 
[23] Torii et al. 1999; 
[24] Finley et al. 1996;
[25] Harrus et al. 1996; 
[26] Halpern et al. 2001; 
[27] BT97;
[28] Murray et al. 2002;
[29] Willingale et al. 2001; 
[30] Marshall, Gotthelf et al. 1998; 
[31] Kaaret et al. 2000;
[32] Pivovaroff et al. 2001;
[33] Hughes et al. 2001;
[34] Camilo et al. 2002;
[35] Marsden et al. 1997; 
[36] Torii et al. 1998; 
[37] Gotthelf et al. 2000;
[38] Mereghetti et al. 2002}
\end{table*}

 
\section{Criteria for the data analysis}
Our aim is to estimate the fraction of the rotational energy loss of a
neutron star going into X-rays.  This excludes from the sample the
X-ray pulsars powered by accretion from binary companions (see
e.g. Bildsten et al. 1997), and the Anomalous X-ray Pulsars
(Mereghetti 2002), the luminosities of which are inconsistent with
rotational energy loss in case of neutron stars. According to these
criteria, the current sample consists of the 41 pulsars listed in
Table~1.
 
Even in the case of spin-down powered neutron stars, some
contribution(s) to the X-ray flux can derive from cooling. Thus we
include only the following components in the calculation of the X-ray
luminosity: {\sl (i)} the non-thermal (pulsed) emission that
originates in the neutron star magnetosphere; {\sl (ii)} the thermal
emission resulting from the re-heating of (parts of) the neutron star
surface by backward accelerated particles; {\sl (iii)} the flux from
the extended synchrotron nebulae powered by the relativistic particles
and/or magnetic fields ejected by the neutron star.  As a rule, when
possible we subtracted the contribution to the X-ray emission from the
extended supernova shells.  At energies $\ga 2$ keV, the contribution
from the neutron star cooling becomes unimportant. Thus, by
considering the energy range 2-10 keV we reduce the uncertainties
related to the subtraction of the cooling components.

\begin{figure*}
\resizebox{13cm}{!}{\includegraphics{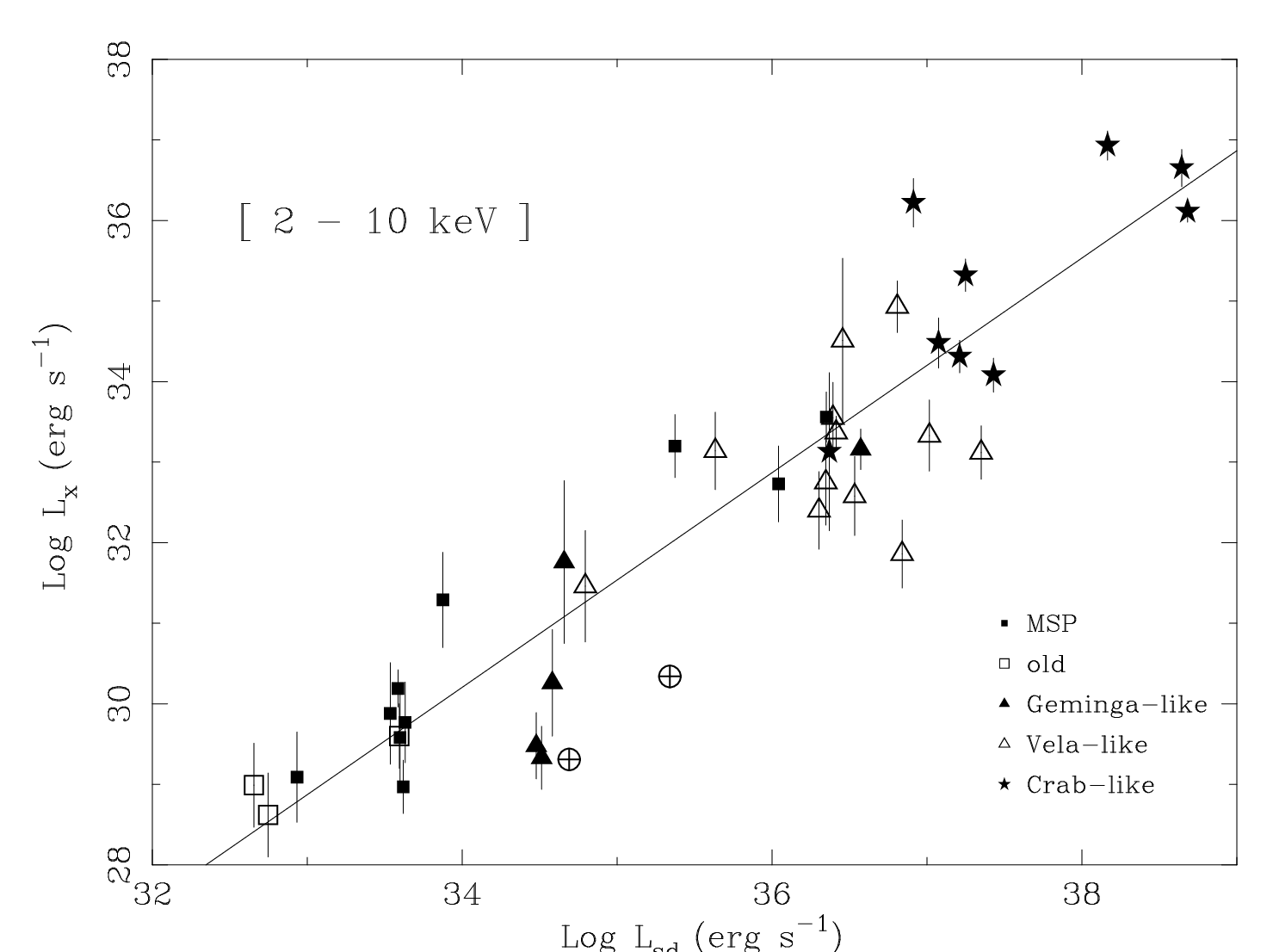}}
\hfill\parbox[b]{45mm} {\caption{The X-ray luminosity in the band
(2-10) keV against the spin-down luminosity for the 41 sources of our
sample.  The objects are grouped in 5 classes and labeled accordingly:
Millisecond Pulsar ({\it filled squares}), old pulsar ({\it empty
squares}), Geminga-like ({\it filled triangles}), Vela-like ({\it
empty triangles}) and Crab-like ({\it filled stars}).  The {\it solid
line} represents our equation [3]; it has been calculated on 39
objects (thus excluding PSR J0114+58 and PSR J0538+2817, also labeled
with a circle filled with a cross and without their huge error bar;
see Table 1).}
\label{fig1}}
\end{figure*}

We reviewed all the data in the literature for each source of our
sample.  When good observations were available in the 2-10 keV band,
we adopted the reported flux. When only fluxes in a different energy
range were available, we converted them to our range, adopting the
parameters of the best spectral fit reported in the literature (see
the references in Table~1).  The luminosities $\lxasca$ are calculated
assuming isotropic emission and the distances listed in Table~1.

As a main improvement with respect to previous similar studies, we
derive a {\it weighted fit} of the $\lx$ versus $\erot$ data.  For
this purpose, we have taken into account the uncertainty associated to
each value of $\lx.$ Contributions to such uncertainties have
different origins:

\begin{itemize}
\item[(a)] {\bf Uncertainties in the distances}; for most of the
pulsars, we have used the distances derived from the dispersion
measure (DM; Pulsar Catalogue {\tt
http:\break //pulsar.princeton.edu/pulsar/catalog}).  They rely on a model
for the electron distribution of the interstellar medium (Taylor \&
Cordes 1993). For individual sources this can lead to an error in the
distance up to a factor $\ga 3$ (e.g. the case of PSR~J1119$-$6127,
Camilo et al. 2000, Crawford et al. 2001), but, when averaged over the
pulsar population, the typical errors reduce significantly (Taylor \&
Cordes 1993). When only the distance inferred from the DM was
available, we conservatively adopted an uncertainty of $\pm40\%$. This
translates in an error $\sim 0.3$ in $\log\lx$. When other
determinations of $d$ exist (e.g. from parallax measurements or from
associations with supernova remnants) we opted for the most accurate
ones, assuming the uncertainties reported for the related measure.
\item[(b)] {\bf Statistical errors in the number of photons}; the
errors due to photon counting statistics dominate the weak pulsars
first detected with {\sl ROSAT}, yielding uncertainties up to $\sim
0.5$ in $\log\lx;$ statistical errors in the count rate are negligible
for brighter objects.
\item[(c)] {\bf Errors depending on interstellar absorption}; the
conversion from the observed flux to that emitted at the source
depends on the amount of interstellar absorption, which is typically
deduced (in a model dependent way) from the X-ray spectral fits.  We
evaluated the errors deriving from the poor knowledge of N$_H$ on a
case by case basis.
\item[(d)] {\bf Errors depending on spectral and spatial modeling};
collecting area, integration time, spectral and spatial resolution
span a large range of values for the observations of different
sources, implying that they have been studied with different degree of
details.  This introduces strong differences in the uncertainties
related to the modeling of each source. So we have studied them
individually.
\end{itemize}
  
In the computation of the error bars we first varied $\nh$ and the
spectral parameters within the uncertainties reported in Table~1,
taking the resulting maximum and minimum fluxes.  The errors on
$f_{\rm X}$ are then propagated accounting for the distance
uncertainties and for the error due to photon counting statistics.

\begin{figure*}
\resizebox{\hsize}{!}{\includegraphics{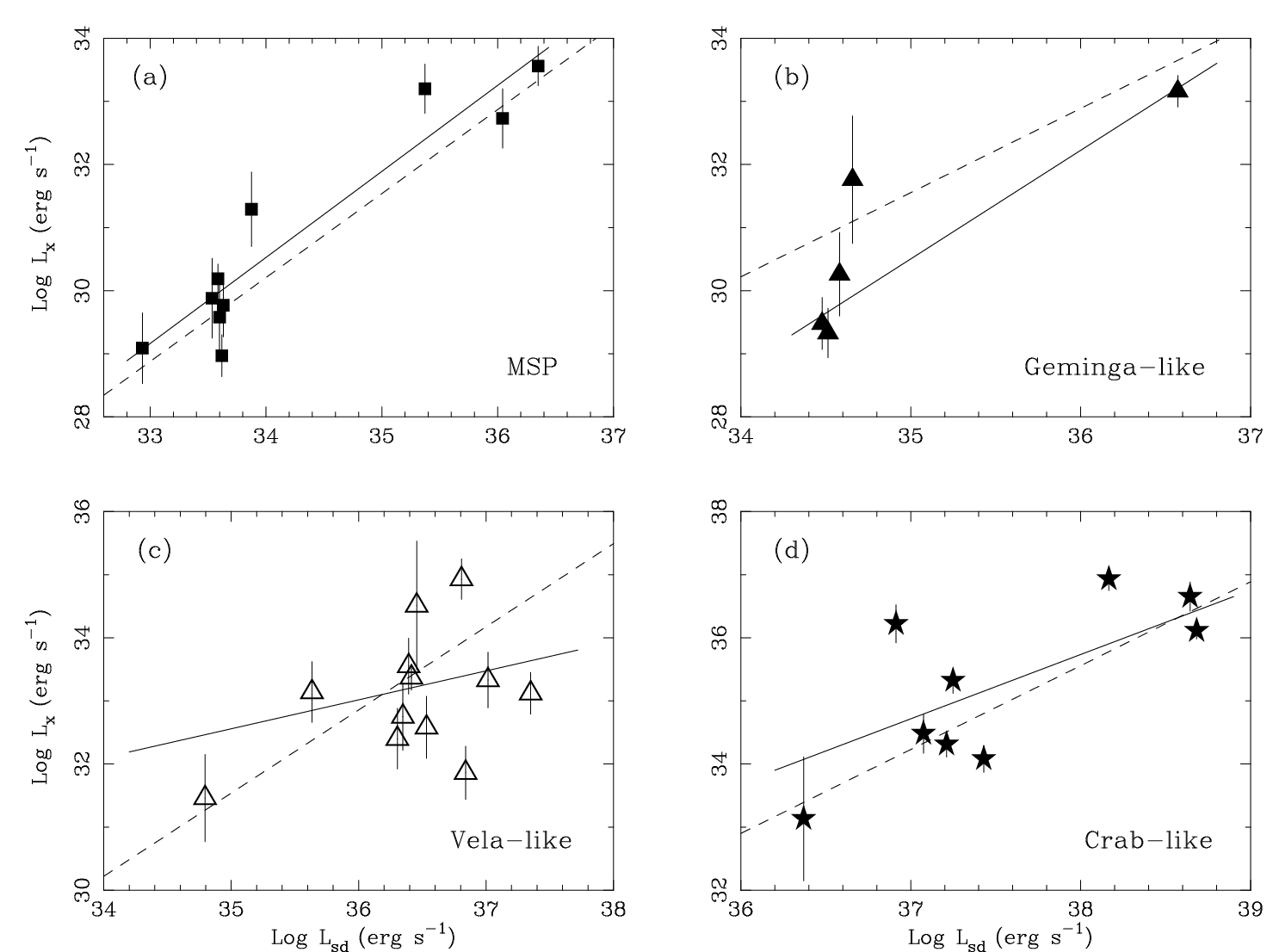}}
\caption{The X-ray luminosity in the (2-10) keV band plotted against the  
spin-down luminosity for 4 subclasses of objects in our sample.
The {\it solid lines} represent the best fits 
of a relation of the type $\log\lxasca=m\log\erot - q$
calculated for the sources of each subclass: they are
$\log\lxasca=(1.38 \pm 0.10)\log\erot - 16.36 \pm 3.64$ 
                            for the Millisecond Pulsars ({\it panel a}), 
$\log\lxasca=(1.71 \pm 0.17)\log\erot - 29.35 \pm 6.15$ 
                            for the Geminga-like sources ({\it panel b}), 
$\log\lxasca=(0.46 \pm 0.23)\log\erot + 16.61 \pm 8.33$ 
                            for the Vela-like sources ({\it panel c}) and 
$\log\lxasca=(1.02 \pm 0.10)\log\erot -  3.14 \pm 3.86$ 
                            for the Crab-like sources ({\it panel d}).
The {\it dashed line} is the fit given in equation [3]: see text for details.}
\label{fig2}
\end{figure*}

\section{The current sample}
The 41 pulsars listed in Table~1 can be naturally grouped into five
classes, based on their intrinsic and observational characteristics.
\begin{itemize}
\item[I] {\it Millisecond Pulsars} ({\it MSPs}): Six among the ten
MSPs in our sample have been detected only by {\sl ROSAT}, and four of
these with a very low photon counting statistics ($<50$ photons).
This does not allow to asses which is their best spectral model.  The
objects studied in more detail with {\sl ROSAT} indicate that
non-thermal fits are preferred (Becker \& Tr\"{u}mper 1999).  However,
recent observations with {\sl Chandra} of the MSP population in the
globular cluster 47 Tucanae (Grindlay et al. 2001), indicate that a
thermal component may be present at energies $\la 1.5$ keV.
Considering that the possible thermal component would contribute only
slightly to the luminosity in our energy range (2-10 keV), we have
adopted for all the MSPs a power law spectrum with photon index
$\alpha_{ph}=2$ (so that the specific energy flux $f_\nu\propto
\nu^{(-\alpha+1)}$).  The list of sources belonging to this group is
reported in the upper section of Table~1.
\item[II] {\it Old Pulsars:} Only three of the non recycled pulsars in
our sample have characteristics ages ($\tau= P/2\dot P$) greater than
$10^6$ yr: PSR~B0823+26, PSR~B0950+08 and PSR~B1929+10.  {\sl ASCA}
observations of the latter two sources were reported by Wang \&
Halpern (1997), but it was later found that the results for
PSR~B0950+08 were affected by the presence of a nearby AGN (see note
in Wang et al. 1998).  The spectrum of PSR~B1929+10 in the {\sl ASCA}
band can be described by thermal emission from a small fraction of the
star surface ($\la 10-30$ m in size).  This is generally interpreted
in terms of re-heating of the polar caps by backward accelerated
particles in the magnetosphere.  Assuming that the same interpretation
is valid for PSR~B0823+26 and PSR~B0950+08, we use for these objects
the flux obtained by scaling their ROSAT count rates to that of
PSR~B1929+10 and adopting large error bars.
\item[III] {\it Geminga-like:} This class contains the middle aged
pulsars ($\tau\sim 10^5$ yr) for which the internal cooling gives a
substantial contribution, at least at energies below $\sim$2 keV.  For
the three brightest objects of this group (Geminga, PSR~B0656+14 and
PSR~B1055$-$52) accurate spectral modeling exist, allowing to estimate
with a good accuracy and subtract the contribution from the cooling to
$\lx$.  For PSR~B1951+32 our adopted $\lx$ derives mainly from the
synchrotron nebula surrounding the pulsar.  The remaining two sources
(PSR~B0114+58, PSR~J0538+2817) are considerably weak preventing a
detailed spectral analysis; therefore these two sources are not
included in the statistical fit.  However, in order to include them in
the graphical representations of Figures~1 and 4, we have assumed a
power-law model, with $\alpha$ as given in Table~1, and a ratio
$\sim10^{-3}$ between the flux in the power-law and the cooling
component, i.e., an average between Geminga and PSR~B1055$-$52. This
could largely underestimate the flux if it is entirely of
magnetospheric origin.
\item[IV] {\it Vela-like:} This is a rather inhomogeneous group of
$\sim 10^4-10^5$ years old pulsars. Some of them are associated to
supernova remnants.  For most of these objects, the main contribution
to $\lxasca$ arises from their synchrotron nebulae, that typically
have power-law spectra with $\alpha_{ph}$ in the range 1.5$-$2.3
(PSR~B0833$-$45, PSR~B1046$-$58, PSR~J1105$-$6107, PSR~B1706$-$44,
PSR~J1811$-$1926, PSR~B1823$-$13, PSR~B1853+01, PSR~J2229+61).  In the
case of PSR~B1757$-$24 the bulk of the emission in the (2--10 keV)
band seems of magnetospheric origin, whereas in the cases of
PSR~J1420$-$6048, PSR~B1800$-$21, PSR~B2334+61 the number of collected
photons is too small for a detailed analysis.
\item[V] {\it Crab-like:} These are the youngest pulsars in our
sample.  They have been directly seen in the hard energy band, and are
usually bright enough for a careful examination of their spectra.
Their emission consists of two components, a pulsed one of
magnetospheric origin, and a second one (unpulsed) from the
synchrotron nebula. The pulsars of this class are listed in the lower
section of Table~1. Only for PSR~J1119$-$6127 and PSR~J1124$-$5916
observation prevents the possibility of disentangling the pulsed from
the unpulsed components: for the former source due to the low photon
statistics of the {\it ASCA} detection, for the latter object due to
the poor time resolution of the adopted {\it Chandra} detector. 
\end{itemize}

\begin{figure*}
\resizebox{\hsize}{!}{\includegraphics{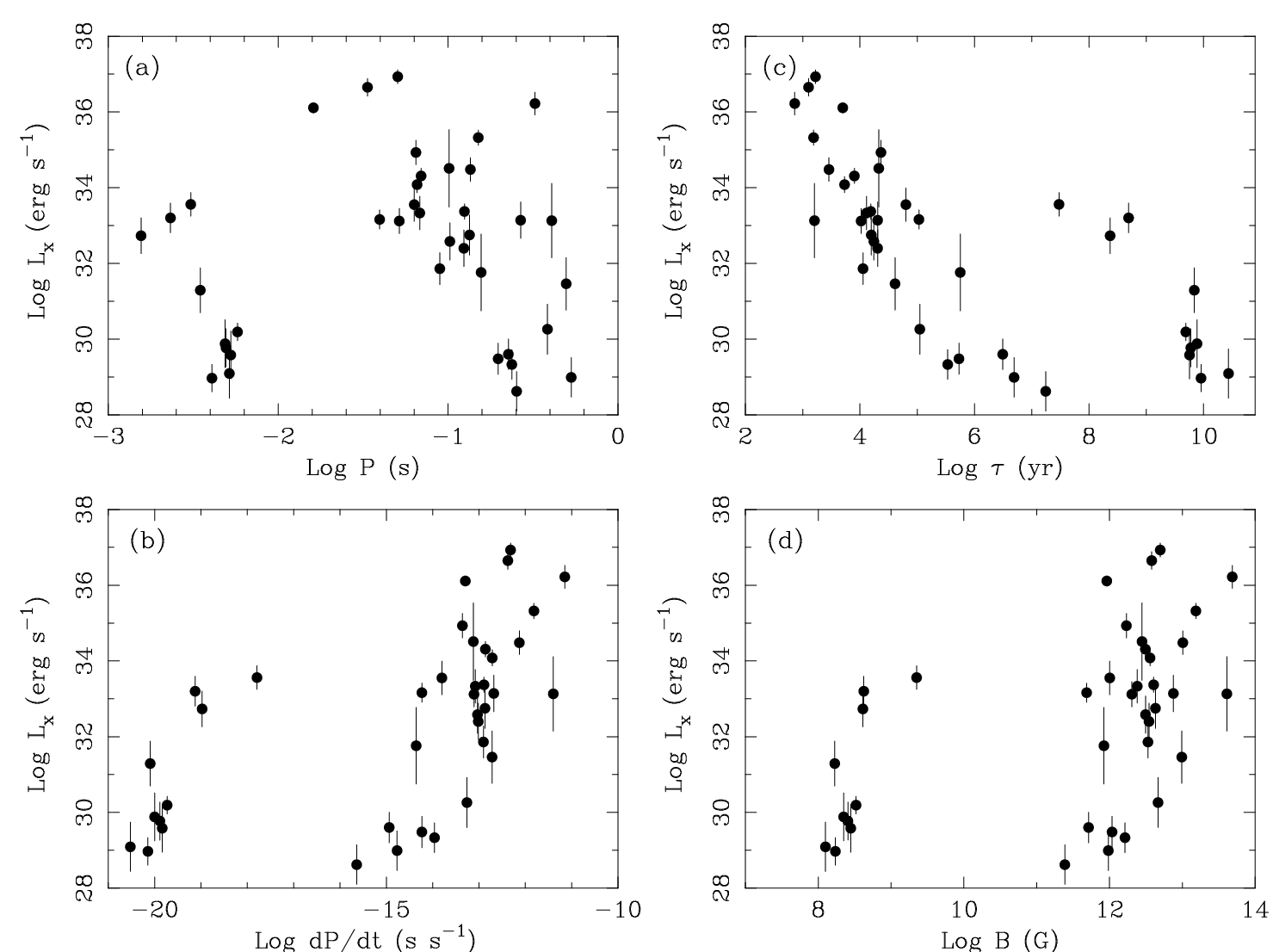}}
\caption{X-ray luminosities in the (2-10) keV band plotted versus the
rotational period ({\it panel a}), the period derivative ({\it panel
b}), the characteristic age ({\it panel c}), and the surface magnetic
field ({\it panel d}).}
\label{fig3}
\end{figure*}

\section{Results}

To facilitate the comparison with previous works, we have first fitted
39 points in Figure~1 with a linear relation, obtaining
\begin{equation}
\log\lxasca~=~1.34~\log\erot~-~15.34
\label{oldfit}
\end{equation}
with a $\chi^2=259.4$ (solid line in Figure~1).  The formal $1\sigma$
intervals of uncertainty are $1.34\pm0.03$ (for the slope) and
$14.36\pm 1.11$ (for the constant term).  Thus the slope $m$ of the
relation (\ref{oldfit}) is slightly flatter than the value of 1.5
derived in the same energy interval for a sample of 15 objects by
Saito (1998), who considered only the pulsed emission.  At lower
energies, {\sl ROSAT} data instead suggest a slope of
$m_{(0.1-2.4)}=1.04\pm 0.09$ for 27 sources (BT97), though a steeper
dependence $m_{(0.1-2.4)}=1.35$ was derived by \"{O}gelman (1995)
using a sample of 7 sources.  At the intermediate energy range 0.2$-$4
keV Seward \& Wang (1988) found a value of 1.39 based on a small
sample of 8 pulsars. We note however that the empirical relation of
$\lx$ with $\erot$ is not fully satisfied as the fit of equation
(\ref{oldfit}) is statistically unacceptable having an extremely large
value of the reduced $\chi^2\sim 7.0$.  The scatter around this
relation is remarkable, even when splitting the data in the different
classes of sources, as shown in Figure~2.

\begin{figure*}
\resizebox{13cm}{!}{\includegraphics{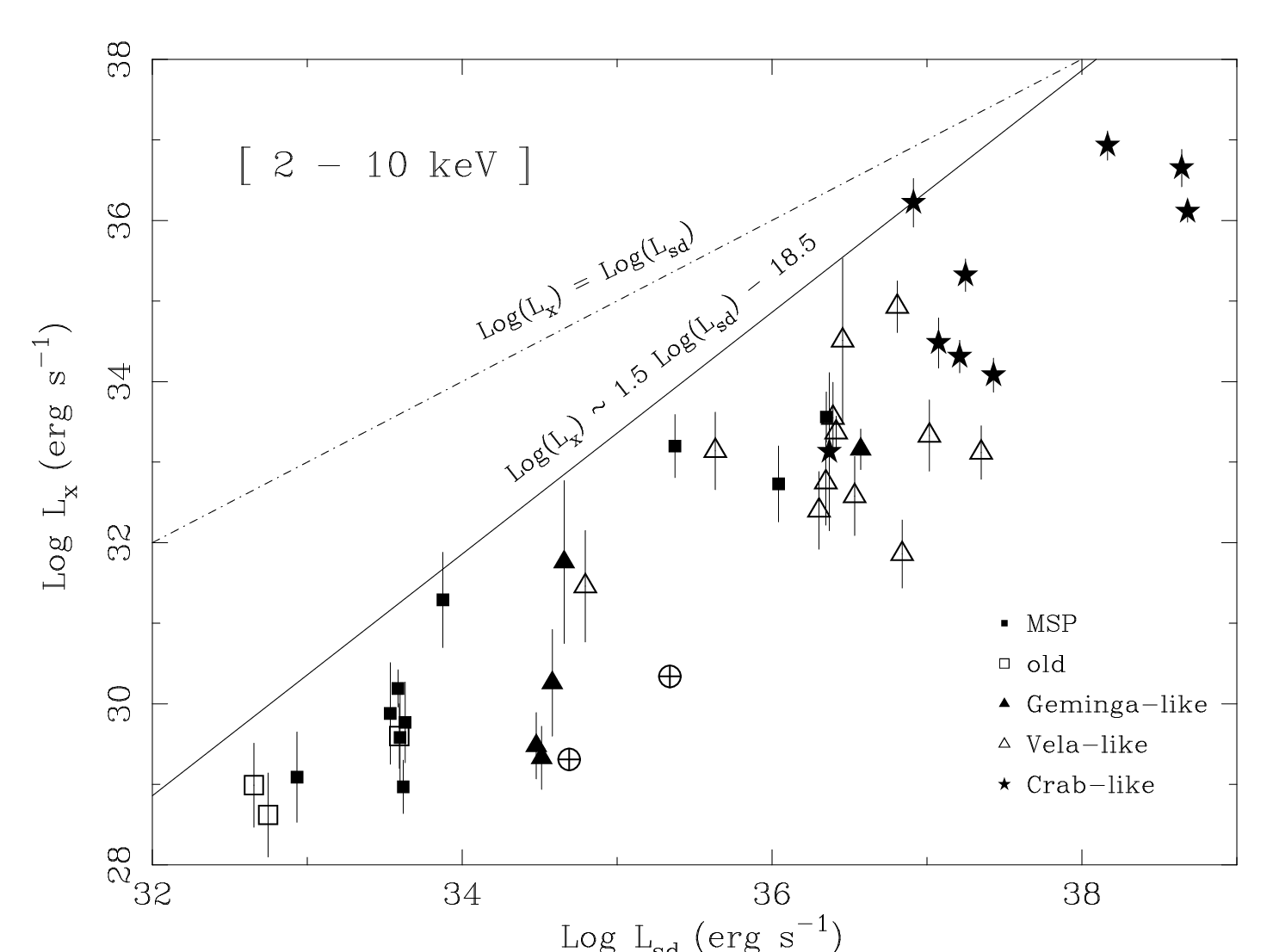}}
\hfill\parbox[b]{45mm}{\caption{The X-ray luminosity in the band 
(2-10) keV against the spin-down luminosity for the 41 sources of our
sample. Objects and labels are as in Figure~1.  The {\it dot-dashed line}
represents the line $\log\lxasca=\log\erot$, whereas the {\it solid
line} is the critical line of equation [6], obtained excluding the two
pulsars PSR J0114+58 and PSR J0538+2817.  All points in this
diagram locate below the solid line.}
\label{fig4}}
\end{figure*}

In the caption of Figure~2, we have reported the results of the linear
fit for 4 distinct subclasses of objects in our sample. We note that
the Crab-like sources (panel d) are evenly distributed on both sides
of the best-fit line of equation (\ref{oldfit}) (dashed line in all
the panels of Figure~2), but their local best fit line (with a slope
of $m_{\rm crab}=1.02$) is flatter than that derived
from fitting all data in our sample. Remarkably, the two sources
located in the Magellanic Clouds (whose relative X-ray luminosity is
not affected by uncertainties in the distance) even show an
anticorrelation between $\erot$ and $\lxasca$: PSR J0537$-$69 (whose
$\erot$ is about 3 times greater than that of PSR B0540$-$69) appears
about ten times dimmer in the {\sl ASCA} band than PSR B0540$-$69: we
note that the ratio between the two luminosities, for these two
objects, scales with the ratio of their period derivatives $\dot P$.

Geminga and the other cooling pulsars have a relatively steep slope
$m_{\rm gem}=1.71$ and (with the exception of PSR~B0355+54) are
underluminous of $1-1.5$ dex relative to the dashed line (eq. [3])
describing the whole sample. This could suggest that either the
(subtracted) thermal component has been overestimated or that these
sources are preferred targets for the mechanisms which could reduce
the luminosity in the 2-10 keV band (see \S 5).

The typical spin down ages, the morphologies and the spectral
characteristics of the Vela-like pulsars classify them between the
Crab-like and the Geminga-like objects.  Thus, it is not surprising
that the data plotted for this group of sources (panel c of Figure~2)
display features of both the Crab-like and the Geminga-like plots. In
fact the slope $m_{\rm vela}=0.46$ is much flatter than that of the
entire sample and a significant number of these objects appear much
dimmer than predicted by equation \ref{oldfit}.

Finally, Figure~2 shows that the MSPs are the sources which better
follow the general $\log\lxasca\,\, {\rm vs} \,\,\log\erot$ relation,
both in terms of slope $m_{\rm msp}=1.38$ and of luminosities.
Interestingly, the MSPs in 47 Tuc observed in the (0.5-2.5) keV
interval with {\it Chandra} appear to have a weaker dependence of
$\lx$ on $\erot$ than those in the field, with a best median estimate
for the slope of $0.55\pm 0.2$ (Grindlay et al. 2002).

\begin{figure*}
\resizebox{13cm}{!}{\includegraphics{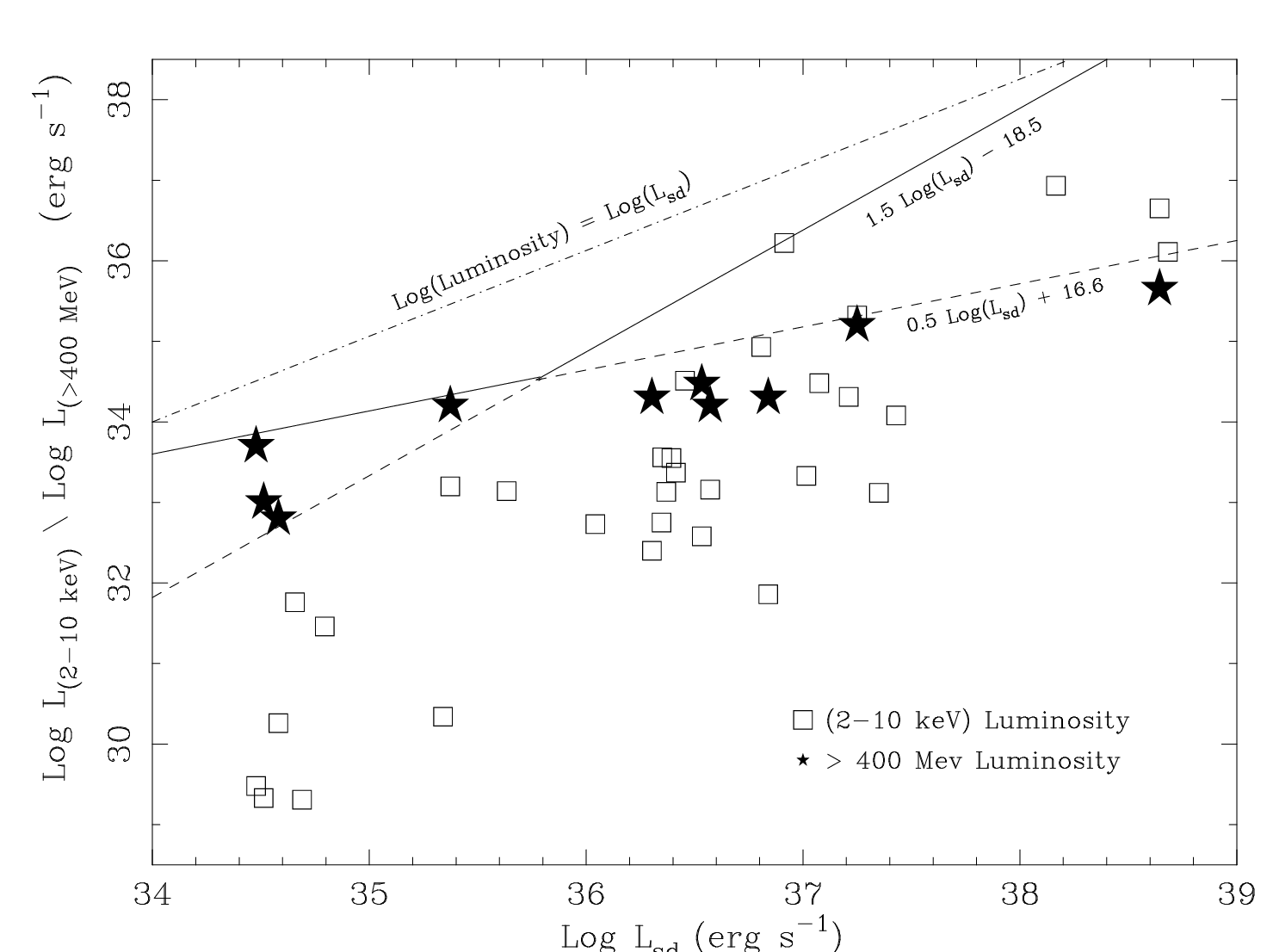}}
\hfill\parbox[b]{45mm}{\caption{High energy luminosity as a function
of $\erot$ in two selected energy domains: in the (2-10) keV ($L_{(2-10) {\rm
keV}}$) and above 400 MeV ($L_{(>400 {\rm MeV})}$).  {\it Stars}
represent young $\gamma$-ray pulsars taken from Zhang \& Harding
(2000; Fig. 3); {\it open squares} refer to the pulsars of our
sample (with the exclusion of MSPs). The {\it dot-dashed line} gives
the locus where luminosity equals the spin-down power. The
{\it dashed lines} gives the critical luminosities (described as
single power laws) in the two energy domains while the broken {\it
solid line} interpolates the two lines, and indicates the observed
change in slope, as discussed in section 5.}
\label{fig5}}
\end{figure*}
 
Figures~3a and 3b illustrate that the scatter in the values of
$\lxasca$ (already appearing in Figure~1) is even larger when the
X-ray luminosities are plotted separately versus the periods and
period derivatives. We also note the lack of any correlation between
the plotted quantities. There are two physical ways of combining $P$
and $\dot P$ to give the characteristic age ($\tau=P/2{\dot P}$) and
the surface magnetic field ($B_s=3.2\times10^{19}\sqrt{P\dot P}$ G) of
a pulsar.  In Figures~3c and 3d we report $\lxasca$ versus these two
quantities: while the spread in the data remains large, a correlation
(already noted by BT97) is visible in $\lxasca$ versus $\tau$, when
the subsample of the MSPs is excluded.

This suggests the possibility to improve the quality of the fit by
using a more general function of $P$ and $\dot P$. In fact, the
proposed physical models for the X-ray emission rely on mechanisms
depending on different combinations of these two quantities.
Therefore, we have explored a fit of the type
\begin{equation}
\log\lxasca~=~a\log P+b\log\dot P+c~.
\end{equation}
Using a $\chi^2$ minimization code, we found $a=-4.00\pm0.13$,
$b=1.34\pm0.03$, $c=47.11\pm0.32$ with a $\chi^2=259.3$, implying that
the fit is still unacceptable. Only enlarging the error bars on
$\log\lxasca$ up to absolutely unreliable values of $\sim 0.9$ (in
units of log(erg s$^{-1}$)), this modeling of the data would become
statistically acceptable (with $\chi^2/{\rm d.o.f.}=1.1$).
Rearranging the best-fit, we get $\log\lxasca=1.34(-2.99\log
P+\log\dot P+c')= 1.34(-3\log P+\log\dot P+c')+0.02\log P$. Since
$\log P$ spans the interval $-3\to 1$, the last addendum is of the
order of the $1\sigma$ uncertainty on the constant term, and thus we
can approximate the formula, getting
\begin{equation}
\log\lxasca~=~1.34\log\erot~-~15.30
\label{newfit}
\end{equation}
which is very close to the best fit of equation (\ref{oldfit}).
This exercise has shown that, when properly accounting for the
uncertainties in the measured $\lxasca$, no combination of $P$ and
$\dot P$ of the type $P^{a}{\dot P}^b$ can fit the data in a
statistically acceptable way.  However, it is remarkable that, among
all the possible combinations of $P$ and $\dot P$, the one which
better describes the data selects again a scaling with $\erot$.

We note that all the data of Figure~4 lie under the critical line 
\begin{equation}
\lxcrit=10^{-18.5}\left  ( {\erot\over \ergs} \right )^{1.48}
\ergs \label{lxcrit}\end{equation}
obtained searching for that line having the minimum {\it weighted}
distance to the observed points. This line gives the maximum
efficiency of conversion of spin-down energy in X-rays:
\begin{equation}
\eta_{\rm x}\equiv {\lxcrit\over \erot}=10^{-18.5}\left (\erot\over
\ergs\right )^{0.48}. \label{etamax}
\end{equation}
For $\erot$ in the interval $10^{32}-10^{38}\ergs$, the efficiency
varies within $10^{-3}<\eta_{\rm x}<0.8.$

\section {Discussion}

$\lxcrit$ can be interpreted as the line giving the maximum luminosity
that can be attributed exclusively to magnetospheric processes, and
the large scatter below that line as due to additional mechanisms that
tend to reduce $\lx$ and that are independent of $P$ and $\dot P$.
Note that the width of the spread in the values of $\lx$ does not vary
over the whole interval of $\erot,$ thus supporting further this
interpretation.

The magnetospheric emission is pulsed and intrinsically anisotropic.
Thus, the detection of a pulsed signal is susceptible to various
geometrical corrections.  In the computation of $\lx$ we have assumed
that the entire beam is seen.  The angle $\alpha$ between the spin
axis and the dipolar magnetic field controls the intrinsic width of
the emission cone. According to Zhang \& Harding (2000), $\lx$ does
depend on $\alpha$.  This effect is particularly pronounced for
$\alpha$=90$^o$ and causes a drop in the luminosity.  Disappointingly,
it is still difficult to have a direct measure of $\alpha$, which is
known, with large uncertainty, only for a handful of sources (Miller
\& Hamilton 1993).  The Vela pulsar is recognized to be an orthogonal
rotator: this can explain the drop of $\lx$ by $\sim 1-2$ orders of
magnitude relative to $\lxcrit$, making the observation consistent
with the prediction, within the error determination.  However, the
small $\alpha \sim 30^o$ of PSR 0656+14 is not sufficient to explain
its displacement from $\lxcrit.$ It is thus plausible that $\alpha$
influences the values of $\lx$ but is not sufficient to explain the
scatter seen in the $\lx$ versus $\erot$ correlation.

Environmental effects can also be a source of scatter for the young
pulsars of the sample. As an example, PSR 0537$-$69 is one order of
magnitude less luminous than the Crab pulsar despite the similarity in
the value of $\erot$ and in the apparent morphology.  Wang et al
(2001) explain the smaller efficiency in terms of a stronger
confinement of the pulsar nebula by a surrounding star formation
region, which combined with the high pulsar velocity ($600$ km
s$^{-1}$) gives a smaller shock radius of a fraction of a parsec. The
relativistic particles responsible for the X-ray emission thus escape
the nebula in a time much shorter than the time of radiative energy
loss.

The 2-10 keV energy band in which the sources are detected is selected
by instrumental needs, but it is not necessarily optimal to sample the
bulk of the broad-band high energy X-ray emission from pulsars.  This
is what emerges from a spectral analysis of a number of bright pulsars
carried out with BeppoSAX data (Massaro et al. 2000).  These new
observations indicate that the photon spectral index ($\alpha$) of the
non-thermal emission is itself a function of energy
$\alpha(E)=a+2b\log(E/E_0)$ (with $a$ and $b$ parameters of the
spectral fit and $E_0=1$ keV) that is consistent with the
observational data when extrapolated in the UV-optical-IR range and at
gamma-ray energies below 30 MeV.  The spectral photon distribution
peaks at a characteristic energy $E_{\rm max}=E_0\cdot 10^{-a/4b}$ th
from source to source.  For Crab, $E_{\rm max}\sim 21$ keV so that the
2-10 keV interval contains a large fraction of $\lx.$ For Vela, the
maximum would fall at about one MeV and then it would imply a low
$\lx,$ just as observed (relative to $\lxcrit$).  The existence of a
characteristic energy or, more generally, of an energy threshold could
cause the deviation of $\lx$ from $\lxcrit.$

The $\gamma$-ray luminosity (above 400 MeV) seems to follow a relation
of the type $L_\gamma\propto \erot^{0.5}$ (Thompson et al. 1997;
Thompson 2001).  In Figure 5 we have drawn (in analogy to $\lxcrit$ in
the 2-10 keV band) the critical $\gamma$-ray luminosity (assuming
emission in one steradian) $\gammacrit \sim
10^{16.6}(\erot/\ergs)^{0.5}\ergs$ derived from Figure 5 of Zhang \&
Harding (2000).  We note that at least three young pulsars (MSPs are
excluded in this discussion), with high $\erot$ lie in between
$\lxcrit$ and $\gammacrit.$ This indicates that for these sources the
X-ray luminosity (in the interval 2-10 keV) dominates over
$L_{\gamma},$ and that the correlation with $\erot$ of the
luminosity integrated above 1 eV is steeper than indicated by Thompson
(2001) above a value of $\erot$ which is around
$10^{36}-10^{36.5}\ergs$ (as indicated in Figure 5).  A possible
interpretation of this steepening is that the physical mechanisms at
play favor the emission in the X-ray channel when $\erot$ exceeds some
value: cascade processes can develop fully so that higher generation
pairs can produce photons at energies of only a few keV.  In this
framework one expects that the slope $n$ of the efficiency which is a
function of $\erot$ and of the selected energy band $\eta_{\bar
\nu}\propto \erot^{n(\bar {\nu})},$ (where $\bar\nu$ is the centroid
frequency of the observational band) gradually increases when moving
toward less energetic frequencies $\bar \nu.$ Since the observed
efficiency in the $\gamma$-ray energy band (above 400 MeV) is
$\eta_{\gamma}\propto \erot^{-0.5}$ while in the 2-10 keV range varies
as $\eta_{\rm x}\propto \erot^{0.5}$, we predict that the maximum
efficiency should become almost independent of $\erot$ (i.e., $n\sim
0$) in an energy band intermediate the two.  Future observations of
pulsars in the energy range of {\it INTEGRAL} between (0.1-10) MeV may
help in testing this prediction.

\section{Conclusions}
The analysis of the current data (with their uncertainties) on the
non-thermal X-ray emission from rotation-powered pulsars shows:
\begin{enumerate}
\item No monomial combination of $P$ and $\dot P$ fits 
$\lxasca$ in a statistically acceptable way.
\item Still, a correlation between the X-ray luminosity
in the band 2$-$10 keV and $\erot$
persists in the data and the preferred scaling is
$\lxasca \propto P^{-4.00}{\dot P}^{1.34} \sim \erot^{1.34}.$  
\item All the data lie below a critical line $\lxcrit \propto \erot^{1.5},$
providing the maximum efficiency of conversion
of $\erot$ in X-ray emission.
\item Geometrical effects, cut-off energy scales and
environment could be the causes of the reduction
in the detected $\lxasca$, thus explaining the large scatter
present in the data.

\end{enumerate}

After comparing our results with those derived similarly in the
$\gamma$-ray band, we suggest that the maximum efficiency of
conversion of spin-down power in high energy emission should be almost
independent of $\erot$ in the energy band that will be explored by
{\it INTEGRAL}.

\begin{acknowledgements}
Part of this work was supported by the Italian ASI ARS 1R272000 grant.
We thank the referee Yoshitaka Saito for very useful and stimulating
comments.
\end{acknowledgements}


\end{document}